\documentclass[12pt]{iopart}
\usepackage{epsfig}
\usepackage{graphicx}
\usepackage{epsf}
\begin{document}

\title[Rapidity and $k_T$ dependence of HBT in Au+Au at 200 GeV with
  PHOBOS] {Rapidity and $k_T$ dependence of HBT correlations in Au+Au
  collisions at 200 GeV with PHOBOS}

\author{Burt Holzman$^1$, for the PHOBOS collaboration:}
\noindent
B~B~Back$^2$,
M~D~Baker$^1$,
M~Ballintijn$^3$,
D~S~Barton$^1$,
R~R~Betts$^4$,
A~A~Bickley$^5$,
R~Bindel$^5$,
A~Budzanowski$^3$,
W~Busza$^3$,
A~Carroll$^1$,
M~P~Decowski$^3$,
E~Garc\'{\i}a$^4$,
N~George$^{1,2}$,
K~Gulbrandsen$^3$,
S~Gushue$^1$,
C~Halliwell$^4$,
J~Hamblen$^6$,
G~A~Heintzelman$^1$,
C~Henderson$^3$,
D~J~Hofman$^4$,
R~S~Hollis$^4$,
R~Ho\l y\'{n}ski$^7$,
B~Holzman$^1$,
A~Iordanova$^4$,
E~Johnson$^6$,
J~L~Kane$^3$,
J~Katzy$^{3,4}$,
N~Khan$^6$,
W~Kucewicz$^4$,
P~Kulinich$^3$,
C~M~Kuo$^8$,
W~T~Lin$^8$,
S~Manly$^6$,
D~McLeod$^4$,
A~C~Mignerey$^5$,
R~Nouicer$^4$,
A~Olszewski$^7$,
R~Pak$^1$,
I~C~Park$^6$,
H~Pernegger$^3$,
C~Reed$^3$,
L~P~Remsberg$^1$,
M~Reuter$^4$,
C~Roland$^3$,
G~Roland$^3$,
L~Rosenberg$^3$,
J~Sagerer$^4$,
P~Sarin$^3$,
P~Sawicki$^7$,
W~Skulski$^6$,
P~Steinberg$^1$,
G~S~F~Stephans$^3$,
A~Sukhanov$^1$,
J~-L~Tang$^8$,
M~B~Tonjes$^5$,
A~Trzupek$^7$,
C~Vale$^3$,
G~J~van~Nieuwenhuizen$^3$,
R~Verdier$^3$,
F~L~H~Wolfs$^6$,
B~Wosiek$^7$,
K~Wo\'{z}niak$^7$,
A~H~Wuosmaa$^2$,
B~Wys\l ouch$^3$

\vspace{3mm}

\small
\noindent
$^1$~Brookhaven National Laboratory, Upton, NY 11973-5000, USA\\
$^2$~Argonne National Laboratory, Argonne, IL 60439-4843, USA\\
$^3$~Massachusetts Institute of Technology, Cambridge, MA 02139-4307, USA\\
$^4$~University of Illinois at Chicago, Chicago, IL 60607-7059, USA\\
$^5$~University of Maryland, College Park, MD 20742, USA\\
$^6$~University of Rochester, Rochester, NY 14627, USA\\
$^7$~Institute of Nuclear Physics, Krak\'{o}w, Poland\\
$^8$~National Central University, Chung-Li, Taiwan

\ead{burt@bnl.gov}

\begin{abstract}
Two-particle correlations of identical charged pion pairs from Au+Au
collisions at $\sqrt{s_{_{NN}}} = 200$ GeV were measured by the PHOBOS
experiment at RHIC. Data for the most central (0--15\%) events were
analyzed with Bertsch-Pratt (BP) and Yano-Koonin-Podgoretskii (YKP)
parameterizations using pairs with rapidities of $0.4 < y < 1.3$ and
transverse momenta $0.1 < k_T < 1.4$ GeV/c.  The Bertsch-Pratt radii
decrease as a function of pair transverse momentum.
The pair rapidity
$Y_{\pi\pi}$ roughly scales with the source rapidity $Y_{\it YKP}$,
indicating strong dynamical correlations.
\end{abstract}

Identical-particle correlation measurements (Hanbury-Brown 
and Twiss, HBT) yield valuable information on the size, shape,
duration, and spatiotemporal evolution of the emission source in 
heavy ion collisions.
Experimentally, the correlation function $C({\mathbf q})$ is defined as
\begin{equation}
\centering
C({\mathbf q}) = \frac{P({\mathbf p}_1, {\mathbf p}_2)}{P({\mathbf p}_1)P({\mathbf p}_2)}
\end{equation}
where $P({\mathbf p}_1, {\mathbf p}_2)$ is the probability of a pair
being detected with relative four-momentum $\mathbf{q} = \mathbf{p}_1
- \mathbf{p}_2$, and $P(\mathbf{p}_1)$ and $P(\mathbf{p}_2)$ are the
single particle probabilities.  The numerator is determined directly
from data, while the denominator is constructed using the standard
event-mixing technique.

The data reported here were collected using the PHOBOS two-arm
magnetic spectrometer during RHIC Run II (2001).  Details of the setup
have been previously described in \cite{phob_NIM}. The spectrometer
arms are each equipped with 16 layers of silicon sensors, providing
charged particle reconstruction both outside and inside a 2 T magnetic
field.  
The primary event trigger was provided by two sets of 16 scintillator
paddle counters, which covered a pseudorapidity range 
$3 < | \eta | < 4.5$. 
Details of event selection and centrality
determination can be found in \cite{phobos1,phobos2}.  The
0--15\% most central events were used in this analysis, equivalent to
$\langle N_{part} \rangle = 310$ as determined
by a Glauber model.

The details of the track reconstruction algorithm can be found in
\cite{phob_spectra200}.
Events with a reconstructed
primary vertex position between -12 cm $< z_{vtx} <$ 10 cm along the
beam direction were selected in order to optimize vertex-finding
precision, track reconstruction efficiency, and momentum resolution.
Only particles which traversed the entire spectrometer were used in
the analysis. A $3\sigma$ cut on the distance of closest
approach with respect to the primary
vertex ($dca_{vtx} < 0.35$ cm) was then applied. 
The final track selection was
based on the $\chi^{2}$ probability of a full track fit, taking into
account multiple scattering and energy loss.
The momentum resolution is $\Delta p/p \sim 1 \%$ after all cuts.
To identify pions, a cut three RMS deviations away from the 
expected mean value of the specific ionization $\langle dE/dx \rangle$ 
for pions was applied.
Contamination from other particle species was studied using HIJING
1.35\cite{HIJING} and a GEANT 3.21 simulation of the full detector.
The contamination from $\textrm{K}^\pm \textrm{K}^\pm$, pp, and
$\overline{\textrm{p}}\,\overline{\textrm{p}}$ 
pairs is less than 1\%; 
non-identical pairs contribute less than 10\% throughout the entire $k_{T}$
range.  To reject ghost pairs, only one shared hit in the weak-field
region and two shared hits in the strong-field region were allowed per
pair.  A two-particle acceptance cut was applied to both data and
background; the criterion for pair acceptance was defined by $\Delta
\phi + 2 \Delta \theta > 0.05$ rad, where $\Delta\phi$ and $\Delta\theta$ are the
relative pair separation in azimuthal and polar angle, respectively.  
About 7.3 million $\pi^+\pi^+$ and 5.5 million $\pi^-\pi^-$ pairs survive all cuts.

Systematic errors were determined by changing two-particle acceptance
cuts, cuts in azimuthal separation, random seeds used in mixed-event
background generation, as well as varying the definition of ``event
class'' to create background events from pairs within narrow and broad
vertex ranges.

Because the event-mixed background is the product of tracks from
different events, it does not {\it a priori} include any multiparticle
correlations.  In order to study the HBT correlation, it is necessary
to apply a weight to account for the Coulomb effect.  
The Coulomb correction can be expressed solely as a function of
relative 4-momentum $q$,
\begin{equation}
F_R(q) = \frac{F_c(q)}{F_{pl}(q)} = \frac{\int d\vec{r}\,
|\psi_c(\vec{r})|^2 S (\vec{r})} {\int d\vec{r}\,
|\psi_{pl}(\vec{r})|^2 S(\vec{r})}
\end{equation}
where $S(\vec{r})$ is the relative separation of the particle pair,
and $\psi_c$ and $\psi_{pl}$ are the Coulomb and plane wave-functions,
respectively.  A closed-form approximation and numerical correction
for this relation was derived in \cite{859_hbt} for $\lambda = 1$.
For a variable $\lambda$, 
\begin{equation}
F_R(q, \lambda) = \frac{(1-\lambda) + \lambda(1 + e^{-q^2R^2}) F_R(q)}
                       {1 + \lambda e^{-q^2R^2} }
\end{equation}

This prescription is nearly equivalent to the corrections
applied by the CERES, STAR, and PHENIX experiments
\cite{ceresCoul,star200,phenix200}; our results showed no significant
change using either correction method.  The method is applied
iteratively, successively fitting distributions of the 
correlation function $C(q)$ and
iteratively applying the fit value $R$ to a new $S(\vec{r})$.
Typically 2 or 3 iterations are sufficient for convergence.

$C({\mathbf{q}})$ is typically fit to a Gaussian source in three
dimensions, the so-called Bertsch-Pratt parameterization \cite{bertschpratt},
\begin{equation}
\label{eq:osl+cross}
C({\mathbf{q}})  =
1 + \lambda e^{-(       
        q_{o}^2 R_{o}^2 + 
        q_{s}^2 R_{s}^2 +
        q_{\ell}^2 R_{\ell}^2 + 
        2 q_{o} q_{\ell} R_{o\ell}^2)}
\end{equation}
The correlation function was also fit to the YKP 
parameterization \cite{yanokoonin}, 
\begin{equation}
\label{eq:ykp}
C({\mathbf{q}})  =
1 + \lambda e^{-(       
        q_{\perp}^2 R_{\perp}^2 + 
        \gamma^2(q_{\|}-\beta q_{\tau})^2 R_{\|}^2+
        \gamma^2(q_{\tau}-\beta q_{\|})^2 R_{\tau}^2         
        )}
\end{equation}
where $\beta$ is the longitudinal velocity of the source and $\gamma =
1/\sqrt{1-\beta^2}$, $q_{\perp}$ and $q_{\|}$ the relative 3-momentum
difference projected in the transverse and longitudinal directions
respectively, and $q_{\tau}$ the relative difference in energy.  
In order to compare with lower energy, the data presented
was fit in the longitudinal co-moving system (LCMS) frame.

\begin{figure} [h]
 \begin{center}
 \epsfysize=3.4in
 \epsfbox{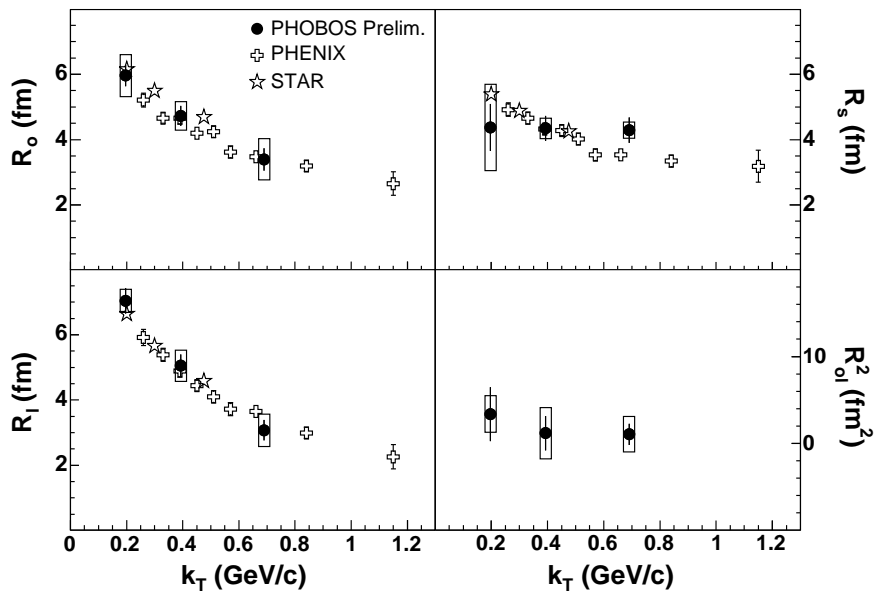}
 \end{center}
\caption{\label{fig:pim_comp}
Bertsch-Pratt radii as a function of pair transverse momentum $k_{T}$ 
for Au+Au at 200 GeV from PHOBOS, 
STAR \cite{star200} and PHENIX \cite{phenix200}. The boxes represent
PHOBOS systematic error.}
\end{figure}
In Fig. \ref{fig:pim_comp}, the Bertsch-Pratt radii are presented as a
function of pair transverse momentum $k_{T}$ for $\pi^{-}\pi^{-}$ pairs.  For
comparison, data from STAR \cite{star200} and PHENIX
\cite{phenix200} at $\sqrt{s_{_{NN}}} = 200$ GeV are also shown.
The PHOBOS data were analyzed in the LCMS frame within the rapidity range
$0.4 < y < 1.3$, while the other data are at mid-rapidity
($-0.5 < y < 0.5$).  The three-dimensional correlation functions were
fit to Eq. (\ref{eq:osl+cross}) using the log-likelihood method.
$R_{s}$ weakly varies as a function of $k_{T}$, while $R_{o}$ and
$R_{\ell}$ decrease rapidly with increasing $k_{T}$.
\begin{figure} [h]
 \begin{center}
 \epsfysize=3.6in
 \epsfbox{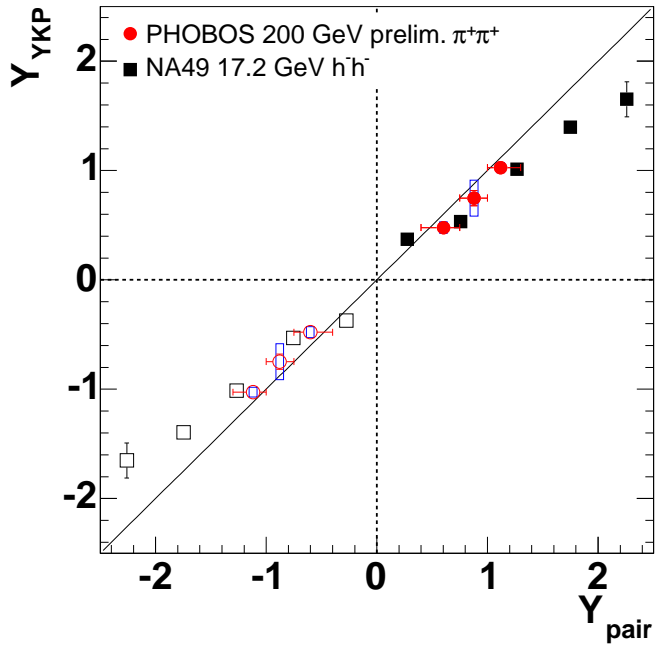}
 \end{center}
\caption{ 
\label{fig:pho_yykp} 
Source rapidity ($Y_{\it YKP}$) as a
function of pair rapidity ($Y_{\pi\pi})$ for PHOBOS (circles) and NA49
(squares) \cite{na49ykp}. 
The line at $Y_{\it YKP} = Y_{\pi\pi}$ is drawn to guide the eye. 
The boxes represent PHOBOS systematic error.}
\end{figure}

In Fig. \ref{fig:pho_yykp}, the extracted value of the source
rapidity $Y_{\it YKP}$ is
plotted as a function of pair rapidity for $\pi^{+}\pi^{+}$ pairs with
$0.1 < k_T < 1.4$ GeV/c.  
The data from NA49 \cite{na49ykp} at lower
energy is also plotted; however, it should be noted the
presented NA49 data covers only $0.1 < k_T < 0.2$ GeV/c.
  The pair rapidity strongly scales
with source rapidity, indicating the presence of strong
position-momentum correlations.  The solid line at $Y_{\it YKP} = Y_{\pi\pi}$ 
represents a class of models including, but not limited to, boost invariance.

In conclusion, we have extracted HBT parameters from Au+Au collisions
at $\sqrt{s_{_{NN}}} = 200$ using two different parameterizations of the
correlation function. The Bertsch-Pratt parameters show good agreement
between three experiments with very different acceptances.  From the
YKP analysis, the pair rapidity scales strongly with the source
rapidity, indicating a source with strong position-momentum correlations.

This work was partially supported by U.S. DOE grants 
DE-AC02-98CH10886,
DE-FG02-93ER40802, 
DE-FC02-94ER40818,  
DE-FG02-94ER40865, 
DE-FG02-99ER41099, and
W-31-109-ENG-38, by U.S. 
NSF grants 9603486, 
0072204,            
and 0245011,        
by Polish KBN grant 2-P03B-10323, and
by NSC of Taiwan under contract NSC 89-2112-M-008-024.


\section*{References}
\begin{thebibliography}{25}
\bibitem{phob_NIM} Back~B~B {\it et al} 2003 \NIM A {\bf 499} 603 
\bibitem{phobos1} Back~B~B {\it et al} 2000 \PRL {\bf 85} 3100
\bibitem{phobos2} Back~B~B {\it et al} 2002 \NP A {\bf 698} 555 
\bibitem{phob_spectra200} Back~B~B {\it et al} 2004 \PL B {\bf 578} 297
\bibitem{HIJING} Gyulassy~M and Wang~X~N 1994 {\it Comput. Phys. Commun.} {\bf 83} 307 
\bibitem{859_hbt} Ahle~L {\it et al} 2002 \PR C {\bf 66} 054906
\bibitem{ceresCoul} Adamova~D {\it et al} 2003 \NP A {\bf 714} 124
\bibitem{star200} Adams~J {\it et al} 2003 {\it Preprint} nucl-ex/0312009
\bibitem{phenix200} Adler~S~S {\it et al} 2004 {\it Preprint} nucl-ex/0401003
\bibitem{bertschpratt} Bertsch~G 1989 \NP A {\bf 498} 173; Pratt~S 1986 \PR D {\bf 33} 1314
\bibitem{yanokoonin} Yano~F and Koonin~S 1978 \PL B {\bf 78} 556; 
Podgoretskii~M 1983 {\it Sov. J. Nucl. Phys.} {\bf 37} 272 
\bibitem{na49ykp} Appelsh\" auser~H {\it et al} 1998 {\it Eur. Phys. J.} C {\bf 2} 661 
\end{thebibliography}
\end{document}